# Ammonia, Water Clouds and Methane Abundances of Giant Exoplanets and Opportunities for Super-Earth Exoplanets[1]

## A Quick Study of Science Return from Direct-Imaging Exoplanet Missions

Renyu Hu[2,3]
Jet Propulsion Laboratory, California Institute of Technology
November 20, 2014

## Executive Summary

Future direct-imaging exoplanet missions such as WFIRST/AFTA, Exo-C, and Exo-S will measure the reflectivity of exoplanets at visible wavelengths. The exoplanets to be observed will be located further away from their parent stars than is Earth from the Sun. These "cold" exoplanets have atmospheric environments conducive for the formation of water and/or ammonia clouds, like Jupiter in the Solar System. We study the science return from direct-imaging exoplanet missions, focusing on the exoplanet atmospheric compositions. First, the study shows that a low-resolution ($R$=70) reflection spectrum of a giant exoplanet at 600 – 1000 nm, for a moderate signal-to-noise ratio of 20, will allow measurements of both the pressure of the uppermost cloud deck and the mixing ratio of methane, if the uppermost cloud deck is located at the pressure level of 0.6 – 1.5 bars. Further increasing the signal-to-noise ratio can improve the measurement range of the cloud deck pressure to 0.2 – 4 bars. The strong and the weak absorption bands of methane allow the simultaneous measurements of cloud and gas; when the uppermost cloud deck is located shallower than the pressure level of 0.2, the weak bands are muted, and the cloud deck pressure and the mixing ratio of methane are not distinguishable from a single reflection spectrum. Second, future direct-imaging exoplanet missions may detect the broadband reflectivity of a few super-Earth exoplanets. If having $H_2O$-dominated atmospheres, directly imaged super Earths are likely to have water clouds located shallower than $10^{-3}$ bars. The very high clouds on these planets would mute most gas absorption features except for $H_2O$, and these planets would occupy a confined phase space in the color-color diagrams. In sum, direct-imaging exoplanet missions may offer the capability to broadly distinguish $H_2$-rich giant exoplanets versus $H_2O$-rich super-Earth exoplanets, and to detect ammonia and/or water clouds and methane gas in their atmospheres.

---

[1] Copyright 2014 California Institute of Technology. U.S. Government sponsorship acknowledged.
[2] Hubble Fellow
[3] Email: renyu.hu@jpl.nasa.gov

# 1 Introduction

The discovery of exoplanets has greatly extended the horizon of planetary exploration. More than a thousand exoplanets have been detected up to now, and among them are gas giant planets, rocky planets, and planets that may have a great amount of water (Howard 2013; Batalha 2014; Marcy et al. 2014). The spectra of short-period giant exoplanets, and several Neptune- and sub-Neptune-sized exoplanets orbiting low-mass stars, have been taken. The spectra reveal the thermal emission of the planets, or the transmission through their atmospheres (Seager & Deming 2010). These measurements have indicated molecular absorptions of $H_2O$, $CO$, $CH_4$, and $CO_2$, and in some cases, the effects of clouds and hazes in the atmospheres (Burrows 2014). The current observations of exoplanet atmospheres using the transit technique work the best for planets close to their parent stars. Due to stellar irradiation, these planets generally have warm and hot atmospheres that are very different from any planetary atmospheres in the Solar System (Burrows et al. 1997; Seager & Sasselov 1998).

Future direct-imaging exoplanet space missions will provide the capability to directly detect exoplanets of nearby stars. The current mission concepts under study are WFIRST/AFTA, Exo-C, and Exo-S (Spergel et al. 2013; Stapelfeldt et al. 2014; Seager et al. 2014). WFIRST/AFTA and Exo-C would use internal coronagraph to suppress the stellar light, and Exo-S would use an external occulter to shade a space telescope from the stellar light. The inner working angles of the missions – the smallest angle at which a planet can be detected – determine that the exoplanets to be observed are sufficiently separated from their parent stars. These exoplanets will have atmospheres much colder and different chemical states than the atmospheres observed currently by the transit technique.

All three missions would measure broadband fluxes and low-resolution spectra of exoplanets at visible wavelengths. The visible-wavelength fluxes from the exoplanets are dominated by reflected stellar light (Marley et al. 1999; Seager et al. 2000; Sudarsky et al. 2000). The direct-imaging missions will therefore characterize "cold" exoplanets by measuring their atmospheric reflectivity.

We anticipate a great diversity in the possible spectral features of the reflected light of exoplanets as a result of clouds and gases in their atmospheres. Rayleigh scattering, molecular absorption, and atmospheric condensates determine the reflection spectra of gaseous exoplanets (Marley et al. 1999). Whether there exists clouds is the primary factor that controls the appearance of an exoplanet at visible wavelengths. Depending on the atmospheric temperature, an exoplanet may or may not have clouds. In particular, assuming an atmospheric elemental abundance the same as the Sun, giant exoplanets may have ammonia, water, or silicate clouds in their atmospheres depending on the orbital distances from their parent stars (Sudarsky 2000, 2003; Burrows et al. 2004). The radiative properties of the clouds are sensitive to the vertical extent of the cloudy layer and the sizes of cloud particles (Ackerman & Marley 2001). The elemental abundance of the atmosphere also affects the formation of the clouds and the spectra (Cahoy et al. 2010). Therefore, reflection spectra of exoplanets contain rich information on the composition, and energetic and dynamic processes of exoplanet atmospheres.

Here we report the prospected science return from the direct-imaging exoplanet missions, summarizing a study commissioned by the NASA Exoplanet Exploration Program. In this study, we focus on the measurement of exoplanet atmospheric compositions enabled by observing the exoplanets in reflection. With a hierarchy of models we develop to simulate the cloud formation and the reflection spectra of direct-imaging exoplanets (Hu & Traub, 2015, in preparation), we provide analysis of one of the key questions: what could we learn about the planets from reflection spectra at a modest spectral resolution? Also, the direct-imaging missions may provide the capability to detect Earth- and super-Earth-sized exoplanets, and measure their reflectivity in a handful of broad wavelength bands. Our analysis includes the consideration of potential opportunities to detect super-Earth exoplanets.

In particular, we address the following three questions:
- Can cloud and gas abundances be measured from a visible-wavelength reflection spectrum?
- What can we learn about the evolution of the planets from the cloud and gas abundance measurements?
- Can we identify super-Earth exoplanets from their broadband reflected fluxes?

In this study, we work with the baseline design of WFIRST/AFTA (Spergel et al. 2013). The prospected exoplanets for spectroscopic analysis are gas giant exoplanets that have equivalent orbital distances[4] of 2 – 4 AU around nearby FGK stars (Traub 2014, personal communication), for broadband analysis several super-Earth exoplanets. We consider the capability of measuring the reflectivity spectra of giant exoplanets at 600 – 1000 nm at a spectral resolution of $R$=70. We also consider the capability of measuring the broadband reflectivity of exoplanets with a bandwidth of ~10%, from 400 to 1000 nm. The findings of this report are also generally applicable to Exo-C and Exo-S, because they are sensitive to similar regimes of planetary parameters as WFIRST/AFTA.

The report is structured as follows. In Section 2 we compare our model calculations with the cloud structure and the reflection spectrum of Jupiter, and summarize the lessons we could learn from the observations of Jupiter. Section 3 outlines the observable quantities of the atmospheres that could be derived from the reflection spectra, and the relationship between the observable quantities and the fundamental quantities that define the planetary atmospheres. Section 4 summarizes the information contained in the prospected reflection spectra, and outlines a strategy to extract the information. Finally, in Section 5 we discuss the use of broadband fluxes in distinguishing $H_2$-rich giant exoplanets and $H_2O$-rich super-Earth exoplanets.

---

[4] An equivalent orbital distance is the orbital distance inverse-scaled by the square root of the luminosity of the parent star as compared to the Sun. This quantity specifies how much the stellar irradiation an exoplanet receives as compared to a planet in the Solar System.

# 2 Learning from Jupiter Observations

The reflection spectra of the gas giant planets in the Solar System between 600 and 1000 nm contain information of the compositions and cloud structures of their atmospheres. For example, the reflection spectra of Jupiter that contain strong, intermediate, and weak methane absorption bands can reject simple models of a single reflective cloud deck but suggest a more complex double-layer cloud structure (Sato & Hansen 1979). Comparing the reflection spectrum from the center and that from the limb determines the vertical extend of the upper cloud layer (Sato & Hansen 1979). With the mixing ratio of methane known from the ratio of the strengths between the $H_2$ quadruple lines and the methane absorption bands, characterization of the cloud structure on Jupiter is also possible at a rather low spectral resolution of ~30. Banfield et al. (1998) use narrow-band images of Jupiter obtained by the Galileo spacecraft to constrain that the upper cloud layer is at 750±200 mbar, and that a haze layer exists above the upper cloud layer (i.e., the upper tropospheric haze). The optical depth of the upper cloud layer is highly varied by location, ranging from 0 to more than 20, and this variation is the controlling factor of the colorful appearance of Jupiter (Banfield et al. 1998; Matcheva et al. 2005; West et al. 2006). The composition of the upper cloud layer is inferred to be ammonia, consistent with the prediction of equilibrium condensation cloud models (Weidenschilling & Lewis 1973).

We have developed a model to calculate the cloud density in the atmospheres of exoplanets that orbit their parent stars at 1 – 10 AU (Hu & Traub, 2015, in preparation). The model extends the classical equilibrium cloud model that has successfully predicted the bulk cloud structure of Jupiter (Weidenschilling & Lewis 1973; Atreya et al. 1999) to the regimes of exoplanets. The model considers water and ammonia as potential condensable species, estimates the particle size for calculating the optical properties of clouds, includes the cloud feedback on the adiabatic lapse rate and the albedo of the planet, and considers partial cloud coverage. We have validated the model by reproducing the temperature-pressure profile and the cloud structure of Jupiter (Figure 1). With the cloud optical properties, we calculate the geometric albedo of the model atmosphere by a delta-Eddington source function algorithm (Toon et al. 1989) and an 8-point Gauss quadrature for the disk average.

In addition to the clouds predicted by the equilibrium condensation cloud model, we also include a layer of photochemical haze for some of the models in the calculation of geometric albedos. We are currently implementing a photochemical scheme to directly simulate the formation of the photochemical haze using the general photochemistry model of Hu et al. (2012, 2013). For now, we include the haze layer in an *ad hoc* way to explore its radiative effects. The haze is assumed to have a constant mixing ratio above the uppermost cloud deck for a vertical span of 2 scale heights. The haze is assumed to be white and has an extinction coefficient the same as ammonia ice. Note that for Jupiter it is unlikely that the haze is made of ammonia ice (West et al. 1989). The mixing ratio of haze is adjusted to obtain the desired total optical depth of the haze layer. Figure 2 compares the model spectra of Jupiter, with and without haze, with an observed disk-average spectrum of the planet (Karkoschka, 1994, 1998).

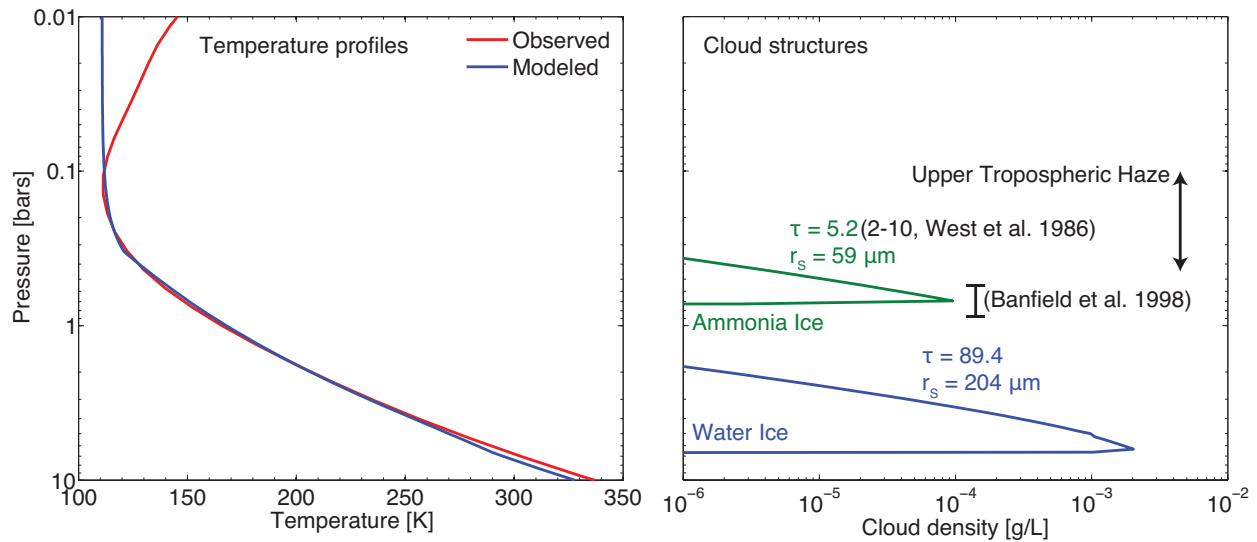

Figure 1: Model of the temperature-pressure profile and the cloud structure of Jupiter. The model is calculatd for a Jupiter-mass, Jupiter-size exoplanet at 5.2 AU of a Sun-like star. The model assumes 3 times solar metallicity, an internal heat flux of 110 K, and an eddy diffusion coefficient of $10^4$ cm$^2$ s$^{-1}$. Varying the eddy diffusion coefficient by 2 orders of magnitude does not lead to appreciable changes to the results. The modeled temperature profile is consistent with the Galileo probe measurements and Cassini CIRS measurements (Seiff et al. 1998; Simon-Miller et al. 2006), except in the stratosphere because aerosol heating is not included in the model. For the cloud structure, the calculated optical depth and mean particle radius (at the cloud base) of each cloud layer are noted on the figure. The model accurately predicts the pressure of the upper cloud layer made of ammonia ice, consistent with the Galileo and Cassini spectral retrieval (Banfield et al. 1998; Matcheva et al. 2005).

While reproducing the cloud structure and the reflection spectrum of Jupiter serves as a validation of the model, it also leads to a number of lessons to keep in mind when analyzing the reflection spectra of gas giant exoplanets.

(1) The equilibrium cloud model can predict the pressures of ammonia and water clouds, and the radiative transfer model using the equilibrium cloud model can predict the strengths of strong methane bands reasonably well. If the temperature profile is correctly calculated, the pressure where the condensation clouds form only depends on the temperature profile and the saturation vapor pressure of the condensable gases. For Jupiter, the model can predict the pressure of the ammonia ice cloud well consistent with observations (Figure 1). The vertical extent of these condensation clouds is small, and most of the optical depth is accumulated near the base of the clouds. Because the column above the uppermost cloud deck controls the strong bands of methane (to be discussed later in Section 4), the equilibrium cloud model can reliably predict the strengths of the strong bands.

(2) The upper tropospheric haze can significantly affect the geometric albedo in weak methane bands. This haze is known to exist in the atmosphere of Jupiter by measuring the ratio between the strong band and the weak band of methane (Sato & Hansen 1979; Banfield et al. 1998). Without this haze, the equilibrium cloud model would overestimate the strength of the methane weak bands. In the atmosphere of Jupiter, this haze is not mainly composed of ammonia ice because ammonia ice features are not detected in infrared spectra (West et al. 1989). The leading candidate is hydrazine ($N_2H_4$) produced from photodissociation of ammonia. Other kinds of photochemical haze may also cause the absorption at wavelengths shorter than 550 nm (Wong et al. 2000, 2003; West et al. 2004).

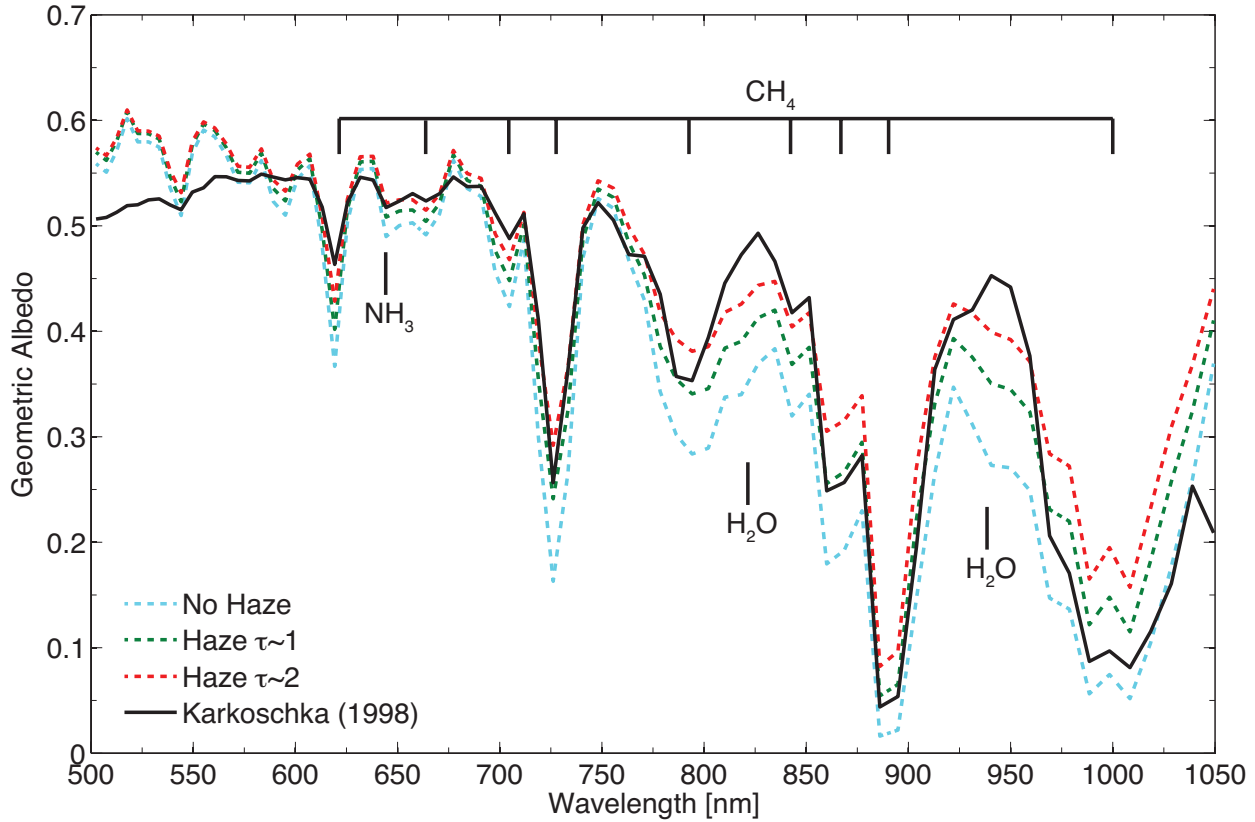

Figure 2: Modeled geometric albedo spectra of Jupiter in comparison with the observed disk-average spectrum (Karkoschka 1998). The model produces a reflection spectrum consistent with Jupiter's in the bulk part. The equilibrium cloud model overestimates the strengths of the methane weak bands, which require adding a diffuse haze layer above the upper cloud layer (Sato & Hansen 1979; West et al. 1989). A thin, purely reflective haze layer, having an optical depth of 1 – 2, can affect the geometric albedo spectrum significantly.

(3) The reflected fluxes outside of major absorption bands are sensitive to trace amounts of absorptive mixtures in the clouds. Considering ideal cases, a fully reflective (i.e., the single scattering albedo=1), isotropically scattering, and infinite atmosphere would have a geometric albedo of 0.69; but the geometric albedo would be only 0.53 for a single scattering albedo of 0.99 (Dlugach & Yanovitskij 1974). The geometric albedo is extremely sensitive to the single scattering albedo when the single scattering albedo is close to unity. The degree of forward scattering of aerosol particles is also very important. For a Henyey-Greenstein phase function with an anisotropy parameter of 0.8, the geometric albedo would become only 0.34 for a single scattering albedo of 0.99. Therefore, The continuum, or color, measures the interplay of the far wings of methane and/or water absorption features, the degree of forward scattering of aerosol particles (mostly controlled by the particle sizes), and a potential sub-unity single scattering albedo of aerosol particles.

(4) In addition to methane bands, minor features of ammonia (650 nm) and water (830 nm and 940 nm) can be seen. There features are more prominent when the atmosphere does not have haze, and an upper tropospheric haze have an optical depth of 1 – 2 would mute the water features, and to a lesser extent, the ammonia feature. These bands may provide opportunities to measure the mixing ratios of water and ammonia.

## 3 Controlling Factors of Exoplanet Reflection Spectra

Learning from the rich history of observing the reflection spectrum of Jupiter, we anticipate that a combination of modest resolution spectra, radiative-transfer spectral analysis, and forward modeling of atmospheric chemistry and cloud formation can lead to exciting discoveries and important insights of the atmospheres of gaseous exoplanets that include Jupiter- and Neptune-sized exoplanets.

Direct-imaging exoplanet missions can measure the reflection spectra of a number of giant exoplanets detected by the radial-velocity measurements of nearby stars (Spergel et al. 2013; Stapelfeldt et al. 2014; Seager et al. 2014). Examples of Jupiter-mass exoplanets that have appropriate angular separations from their parent stars and that would have high enough contrast to allow spectroscopic observations by WFIRST/AFTA are (their equivalent semi-major axis noted in parentheses): HD 114613 b (2.6 AU), Ups And e (2.8 AU), 47 Uma c (2.8 AU), HD 190360 b (3.6 AU), and HD 160691 e (3.8 AU). All of these planets have an equivalent semi-major axis between 2 and 4 AU. In this quick study, we select Ups And e (2.8 AU) and HD 160691 e (3.8 AU) as two examples to represent the group of exoplanets to be characterized by WFIRST/AFTA.

The controlling factors of a gaseous exoplanet's reflection spectrum are its atmospheric molecular compositions and cloud properties. Therefore, these quantities may be directly derived from the spectrum. The question is then, are these independent parameters?

In principle, the cloud properties and the molecular compositions are not independent quantities, because they depend on the same set of fundamental quantities (Figure 3). The atmospheric elemental abundance and temperature determine the molecular composition, balancing the vertical transport, chemical reactions, and photochemical reactions (Line et al. 2010; Moses et al. 2011; Moses et al. 2013; Hu et al. 2014; Venot et al. 2014). The molecular composition and the temperature of the atmosphere determine the cloud structures: when the partial pressure of a certain gas exceeds its saturation vapor pressure, it condenses out to form clouds (Weidenschilling & Lewis 1973; Atreya et al. 1999, Ackerman & Marley 2001).

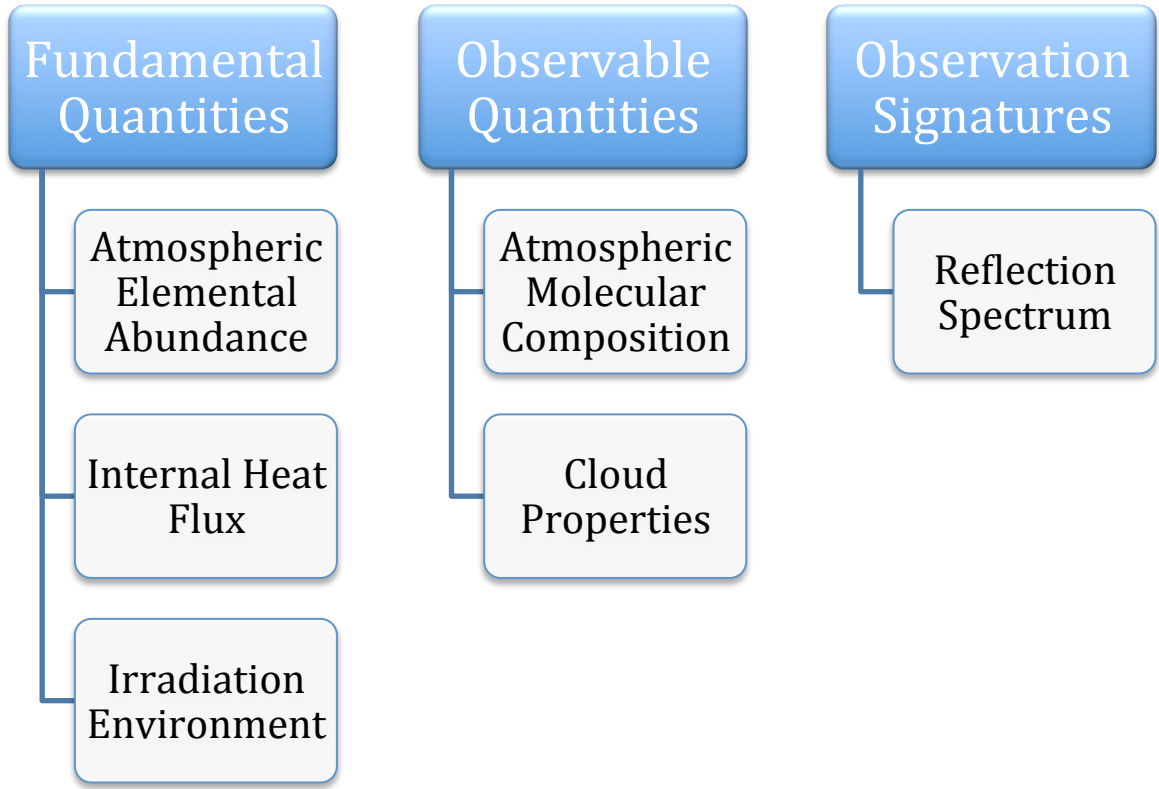

Figure 3: Physical quantities that determine the reflection spectrum of a gaseous exoplanet. The fundamental quantities that determine a planet's appearance are its atmospheric elemental abundances, its internal heat flux, and the amount of irradiation it receives. The internal heat flux is tied to the evolutionary history of the planet. These factors decide presence or absence of clouds in the atmosphere, and the molecular composition of the atmosphere, which in turn determine the reflection spectrum.

Due to stellar irradiation, the existence and types of clouds are strong function of a planet's semi-major axis (Sudarsky et al. 2000, 2003). The planets to be characterized by WFIRST/AFTA, having an equivalent semi-major axis between 2 and 4 AU, are expected to have water and/or ammonia clouds in their atmospheres as Jupiter.

Here we further show that the pressure where the clouds form also depends on both the abundances of water and ammonia gas in the atmosphere (i.e., the metallicity), and the internal heat flux of the planet. Figure 4 shows the cloud top pressure, defined as the pressure at which the vertical optical depth of cloud particles equals to unity, as a function of the atmospheric metallicity for Earth-like, intermediate, and Jupiter-like internal heat fluxes. The general trend is that when the planet is closer to the host star, the cloud top is at a lower pressure; and when the atmosphere is more metal-rich, the cloud top is at a lower pressure. Another significant trend is that when the planet has a greater internal heat flux, the cloud top is at a lower pressure. This picture is complicated by that two potential condensable species, water and ammonia, are present in the atmosphere. For the semi-major axes that AFTA will be sensitive to, ammonia clouds are the uppermost cloud layer, like Jupiter, when the planet has a moderate to high internal heat flux. However, if the planet has a very low internal heat flux, for example for very old planets, we find that the water cloud becomes the uppermost cloud layer. This occurs because the water cloud is in the form of liquid water droplets in this case, and the liquid water droplets absorb ammonia in the atmosphere, thereby deplete ammonia gas above the water cloud, and prevent the formation of high ammonia clouds. For a planet at 2.8 AU, an optically thick ammonia cloud may not be formed even at high internal heat flux because of the high temperature of the atmosphere (right side of Figure 4.). These findings will be reported soon after in a research paper (Hu & Traub, 2015, in preparation).

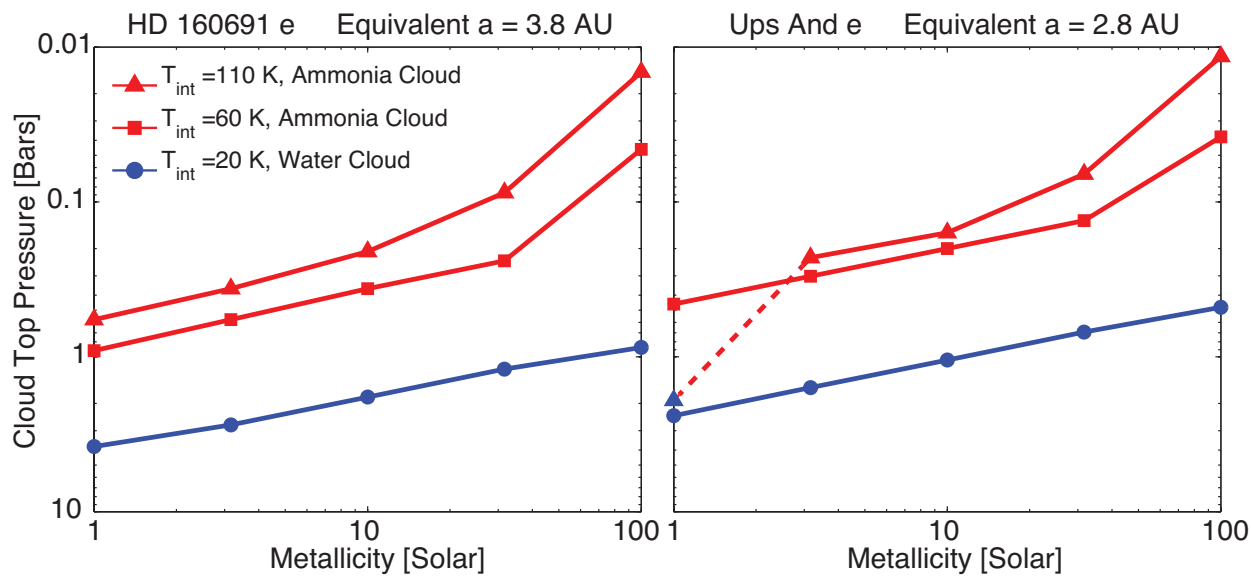

Figure 4: Modeled cloud top pressure for 2 Jupiter-mass exoplanets as a function of the atmospheric metallicity and the internal heat flux. The planets are radial-velocity planets that can be characterized by WFIRST/AFTA. The cloud top pressure is defined to be the pressure where the aerosol vertical optical depth equals to unity.

These results demonstrate that the cloud top pressure does not solely depend on the molecular composition or the semi-major axis, and the internal heat flux may also affect the cloud top pressure. Therefore, determining the cloud top pressure will provide a mean to infer the internal heat flux of a giant exoplanet. The internal heat flux is a key prediction of evolutionary models of giant planets (Baraffe et al. 2003, 2008; Fortney et al. 2006), but its direct measurement would require wide wavelength coverage. The direct-imaging reflection spectroscopy may offer an alternate way to study the evolution of giant exoplanets.

Another implication of these results is that there is no one-to-one relationship between the atmospheric molecular composition and the cloud properties for a given planet. Even with one-dimensional models, other factors including the internal heat flux, may affect the cloud top pressure. In reality, the clouds in the atmosphere of a giant exoplanet may be much more complex than what is shown by the one-dimensional equilibrium cloud model. Even for Jupiter, there are still important unknowns regarding the dynamic processes that drive distinct cloud structures between belts and zones (Ingersoll et al. 2004). Therefore, the cloud structures and the molecular compositions should generally be treated as independent variables when searching a fit to the reflection spectrum. Based on the observable quantities, forward models of atmospheric chemistry, dynamics, and cloud microphysics can seek understanding of the fundamental quantities.

## 4 Information in the Reflection Spectra

Both the cloud top pressure and the mixing ratio of methane, and other species to a lesser extent, determine the spectral shapes of the exoplanets to be characterized by WFIRST/AFTA. Figure 5 shows three series of models computed for a cloud top pressure of 2 bars, 0.4 bars, and 0.1 bars. The 2-bar scenario corresponds to HD 160691 e having a 10x solar abundance atmosphere and an Earth-like (small) internal heat flux, or Ups And e having a solar abundance atmosphere. The 0.4-bar and 0.1-bar scenarios correspond to HD 160691 e having an atmosphere 3 times and 30 times more metal-rich than the solar atmosphere and an internal heat flux the same as Jupiter, or Ups And e having a lower abundance atmosphere (Figure 4).

The strengths of the methane absorption features in the geometric albedo spectra are highly sensitive to the cloud top pressure. When the uppermost cloud deck is shallower than ~1 bars (e.g., $NH_3$ clouds for exoplanets at 2 – 4 AU), the methane absorption features are prominent at 700 – 1000 nm when the mixing ratio of methane is higher than $10^{-4}$. Otherwise, water and ammonia features would dominate the spectrum. When the uppermost cloud deck is deeper than ~1 bars (e.g., $H_2O$ clouds for exoplanets at 2 – 4 AU), the planet would be very dark at all wavelengths longer than 750 nm, and the methane absorption features are prominent at 600 – 750 nm (Figure 5). Comparing the three panels, we see that a deeper cloud would provide a lower baseline continuum for the development of methane features. This is because of the sensitivity to the single scattering albedo; a deeper cloud means that the single scattering albedo of the reflecting layer of the atmosphere is more significantly less than unity.

The question is then whether we could measure the cloud top pressure and the mixing ratio of methane simultaneously from a single reflection spectrum. We have simulated reflection spectra for methane mixing ratios ranging from $10^{-5}$ to $10^{-2}$, and cloud top pressures ranging from 4 to 0.01 bars. The density of sampling is a half order of magnitude for the mixing ratio of methane, and a quarter order of magnitude for the cloud top pressure. For each scenario, we calculate the difference between the scenario and all other scenarios in their reflection spectra, quantified by $\chi^2/dof$ for an assumed signal-to-noise ratio at 600 – 1000 nm. This quantity thus shows the detectability of each scenario, and is illustrated in Figure 6. We explore the nominal signal-to-noise ratio for WFIRST/AFTA (SNR=20), and a very high signal-to-noise ratio (SNR=100). The noise is calculated with respect to a geometric albedo of 0.5, invariable over the wavelength range, assuming that a floor of noise composed of zodiacal light and other systematics, rather than the shot noise of the planetary light, dominates the noise.

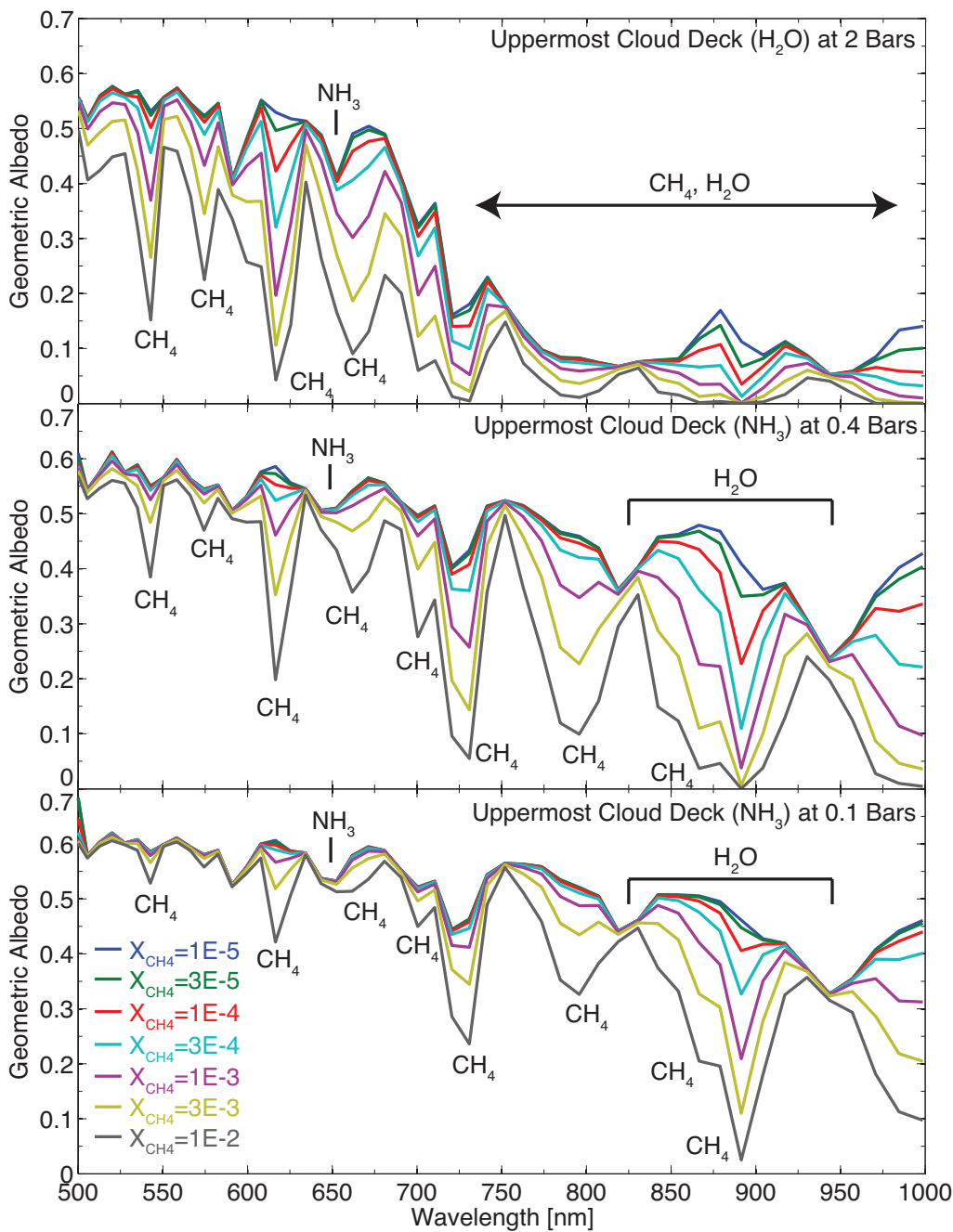

Figure 5: Modeled geometric albedo spectrum of a gaseous giant exoplanet for cloud top pressures ranging from 0.1 to 2 bars, and methane mixing ratios ranging from $10^{-5}$ to $10^{-2}$. The spectra are calculated at a high spectral resolution and binned down to a resolution of $R$=70.

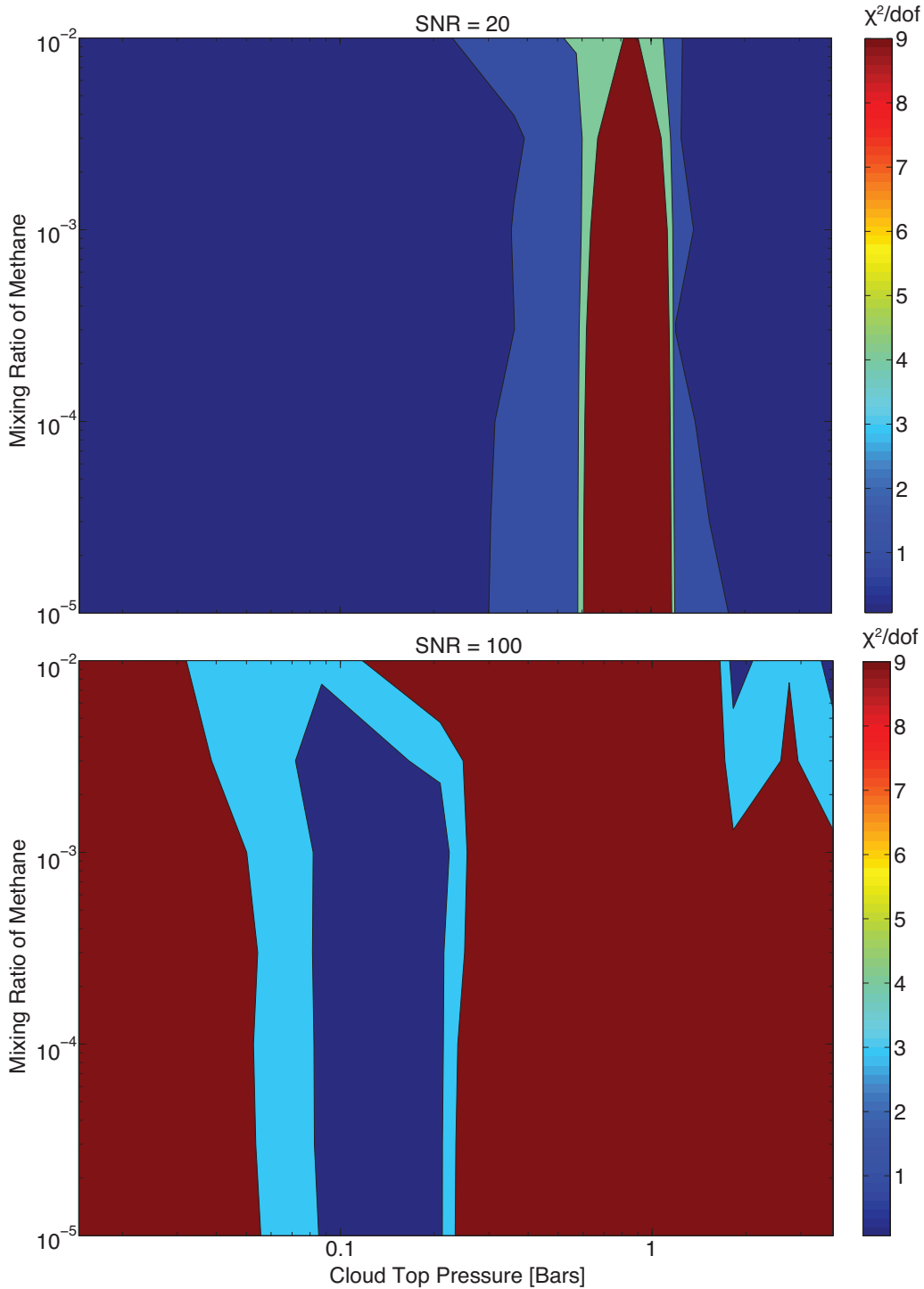

Figure 6: Detectability of the mixing ratio of methane and the cloud top pressure by a single reflection spectrum. The color contours show the minimum $\chi^2$, for two representative signal-to-noise ratios, between a scenario and all other scenarios modeled. With $R$=70, there are 37 spectral elements and dof = 35. A $\chi^2$ greater than 9 means that the mixing ratio of methane can be determined to a precision of a half order of magnitude and the cloud top pressure can be determined to a precision of a quarter order of magnitude, at 3-sigma confidence level.

We find that when the cloud top pressure is greater than 0.2 bars, the cloud top pressure and the mixing ratio of methane can be derived unambiguously from a single reflection spectrum, if the SNR is sufficient. In particular, for the cloud top pressure ranging between 0.6 and 1.5 bars, the two parameters can be derived from a spectrum with the nominal spectral capability provided by WFIRST/AFTA (i.e., 600 – 1000 nm, $R$=70, and SNR=20). In this "sweet spot", the AFTA spectrum will determine the mixing ratio of methane to be more precise than a half order of magnitude, and the cloud top pressure to be more precise than a quarter order of magnitude.

The favorable parameter space for measuring the cloud top pressure and the mixing ratio of methane indeed covers a dominant range for AFTA exoplanets having moderate and high internal heat fluxes, and close-to-solar abundance atmospheres. The moderate and high internal heat fluxes are likely for Jupiter-mass exoplanets according to the planet formation and evolution theories (e.g., Baraffe et al. 2003, 2008). Furthermore, the planets with an equivalent semi-major axis of 3 – 4 AU (such as HD 160691 e) present a better characterization opportunity than the planets with an equivalent semi-major axis of 2 – 3 AU (such as Ups And e), in terms of uniquely measuring the cloud top pressure. Ups And e would likely have an ammonia ice cloud shallower than 0.6 bars, thereby putting the planet outside the identified sweet spot.

Importantly, we find that uniquely determining the cloud top pressure and the mixing ratio of methane from the geometric albedo spectrum would not be possible if the cloud is high at the pressure less than 0.2 bars, even at very high signal to noise ratios (Figure 6). At the spectral resolution of WFIRST/AFTA and other exoplanet direct-imaging missions, fundamental degeneracy exists in this case (see Figure 7 for an example). This parameter space where the cloud pressure and the mixing ratio of methane are degenerate corresponds to planets with highly metal-rich atmospheres (>30x solar abundances) and a Jupiter-like internal heat flux.

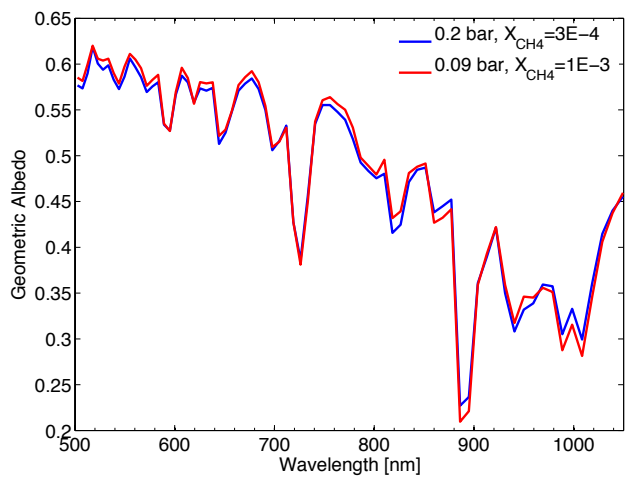

Figure 7: An example of two atmospheric scenarios giving the same reflection spectra. The two cases have different cloud top pressures and different mixing ratios of methane. However, their reflection spectra would be the same at 600 – 1000 nm, and not distinguishable from each other at any reasonable signal-to-noise ratios.

The reason for the "sweet spot" at the cloud top pressure of 0.6 – 1.5 bars and the fundamental degeneracy at the cloud top pressure <0.2 bars is the relative strengths between the strong bands and the weak bands of methane (Figure 8). When the cloud top pressure is shallower than ~1 bars (e.g., $NH_3$ ice clouds), the strong bands are only sensitive to the column of methane above the cloud, but the weak bands depend on the methane column differently and may be use to break the degeneracy. However, when the cloud top is higher than 0.2 bars, the weak bands are not sufficiently deep and no longer offer the diagnostic power. When the cloud top pressure is deeper than ~1.5 bars, the continuum is very low and the methane absorption features do not prominently show up in the spectrum; high signal-to-noise ratios are thus required in this regime.

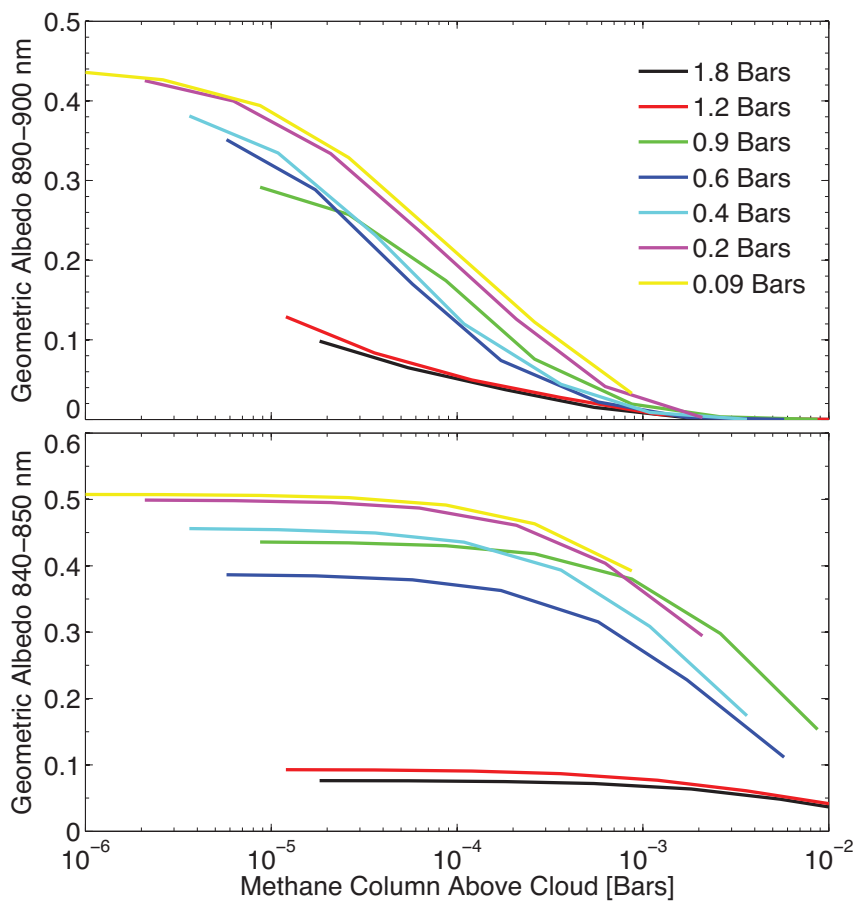

Figure 8: Albedo in a strong band (890–900 nm) and a weak band (840–850 nm) of methane, for the cloud top pressure ranging from 0.1 to 2 bars. The horizontal axis is the product of the methane mixing ratio and the cloud top pressure, i.e., the column of methane above the cloud.

To summarize, when the uppermost cloud deck is deeper than ~0.2 bar, the weak bands and strong bands of methane allow measurement of the methane mixing ratio and the cloud top pressure with one observation of the reflection spectrum (with the caveat of a potential photochemical haze, see below). In this regime, the scenarios with a cloud deck at 0.6 – 1.5 bars can be well studied by a low-resolution reflection spectrum with a signal-to-noise ratio of 20. When the uppermost cloud deck is shallower than ~0.2 bar, the methane mixing ratio and the cloud top pressure are degenerate; more than one spectra at different orbital phases may be required to break this degeneracy. These findings will be reported soon after in a research paper (Hu & Traub, 2015, in preparation).

Once the cloud pressure and the mixing ratio of methane are derived from the reflection spectrum, one can use atmospheric chemistry and cloud formation models to further constrain the metallicity of the planet's atmosphere and the planet's internal heat flux. With the temperature profile calculated, the composition of the clouds may be determined by comparing the measured cloud top pressure with the saturation vapor pressure of candidate species (e.g., $NH_3$ versus $H_2O$). In fact, we expect little ambiguity between $NH_3$ clouds and $H_2O$ clouds once the cloud top pressure is measured, because the pressures of these two types of clouds generally do not overlap for a given planet (Figure 4).

An important cautionary note is that the determination of the cloud top pressure may be complicated by haze or particle cloud coverage. First, a potential photochemical haze would bias the inferred cloud top pressure if not considered. The effect of the photochemical haze is reducing the strengths of the weak bands of methane (Figure 2), which makes the cloud top pressure appear to be higher. There is no clear way to remove the effect of the photochemical haze, because the haze would likely lack any specific spectral features, like the upper tropospheric haze of Jupiter. The investigation of the effect of a potential haze layer would likely rely on atmospheric photochemistry models (e.g., Yung & Strobel 1980; Gladstone et al. 1996; Wong et al. 2000, 2003; Hu et al. 2012, 2013). Second, the inference based on one-dimensional models would indicate a cloud top pressure in average. If the exoplanet has a banded cloud structure like Jupiter, the cloud top pressure derived would likely indicate a weighted average of that of the "belts" and that of the "zones". Because the zones would be brighter and contribute more reflected light than the belts, the one-dimensional model would likely find a value close to the cloud top pressure of the zones corresponding to the updraft portions of convective cells (Ingersoll et al. 2004). It would be intricate to derive a banded cloud structure from a disk-average reflection spectrum.

## 5 Color-Color Diagram and Opportunities for Super-Earth Exoplanets

In addition to measuring spectra of selected radial-velocity giant exoplanets, WFIRST/AFTA can also perform blind searches for exoplanets, in a few broad bands with ~10% band widths at 400 – 1000 nm. The searches may result in the detection of a few exoplanets in the super-Earth and mini-Neptune regimes (Spergel et al. 2013). Exo-S, using the external occulter (starshade), would also be sensitive to super-Earth exoplanets, even Earth-sized exoplanets around nearby stars (Seager et al. 2014). The odds of direct detection of super-Earth exoplanets are corroborated by the *Kepler* result that planets having radii between the radius of Earth and the radius of Neptune, although not existing in the Solar System, are ubiquitous in our interstellar neighborhood and are far more populous than Jupiter-sized exoplanets (e.g., Howard 2013; Batalha 2014).

Super-Earth exoplanets may have massive, non-$H_2$-dominated atmospheres. Super Earths and mini Neptunes obtain their atmospheres by capture from the nebula, degassing during accretion, and degassing from tectonic processes (Elkins-Tanton & Seager 2008). If starting out as an $H_2$-dominated atmosphere accreted from the planet-forming nebula, the atmosphere on a super Earth/mini Neptune can evolve to become a non-$H_2$-dominated atmosphere. Hu & Seager (2014) presented a general classification of the atmospheres on super-Earth exoplanets, which include $H_2$-rich atmospheres, $H_2O$-rich atmospheres, hydrocarbon-rich atmospheres, and oxygen-rich atmospheres. When a super-Earth exoplanet has an $H_2$-rich atmosphere, it would have the same geometric albedo as a giant exoplanet with the same atmospheric composition and internal heat flux. For that case, the results of the previous section apply. Here, we consider an addition case for super-Earth exoplanets: $H_2O$-rich atmospheres.

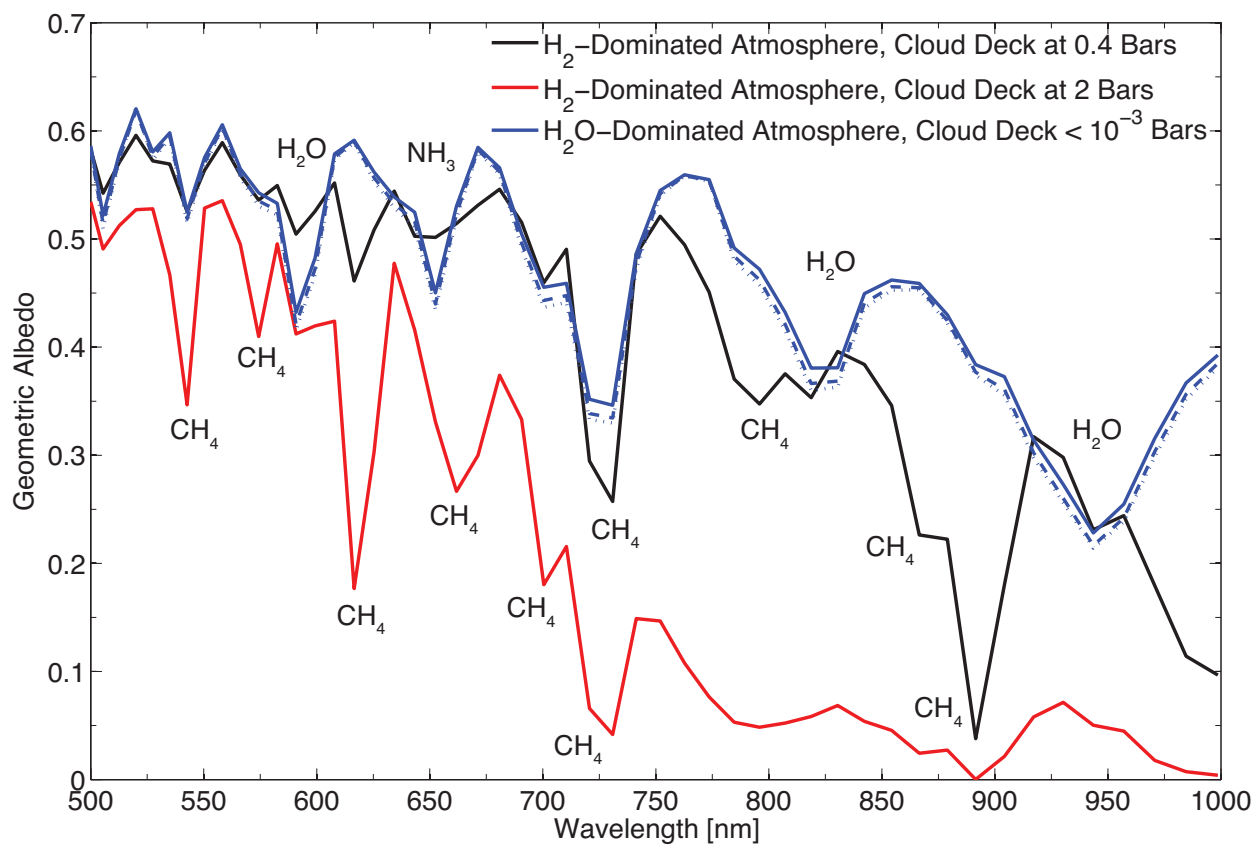

Figure 9: Modeled reflection spectra of super-Earth exoplanets having water-rich atmospheres, in comparison with the reflection spectra of giant exoplanets. The semi-major axis for all models shown is 3.8 AU. The blue lines show the spectra of water-rich atmospheres, for a $H_2O/H_2$ ratio of 1 (solid), 4 (dashed), and 20 (dotted). The mixing ratio of $NH_3$ is scaled with $H_2O$, and the mixing ratio of $CH_4$ is assumed to be $10^{-3}$. The black and red lines show two examples of $H_2$-dominated atmospheres for comparison, also assuming a mixing ratio of $CH_4$ of $10^{-3}$. Because of high water clouds, super-Earth exoplanets having water-dominated atmospheres would be bright, and have water absorption features. The spectra are calculated at a high spectral resolution and binned down to a resolution of $R$=70.

We have modeled the reflection spectra of super-Earth exoplanets having an atmospheric $H_2O$ versus $H_2$ ratio ranging from 1 to 20, at a semi-major axis of 2.8 AU and 3.8 AU. We find that for a wide range of internal heat flux ranging from Earth's to Jupiter's, the modeled exoplanet would have a very high water cloud, with the cloud top shallower than $10^{-3}$ Bars (Hu 2015, in preparation). Equilibrium chemistry and disequilibrium chemistry models have shown that the abundance of methane should be low in a water-rich atmosphere, because methane tends to react with water to form carbon monoxide (Moses et al. 2013; Hu & Seager 2014). The high cloud top and the low abundance of methane determine that the reflection spectrum of a water-rich super Earth would show little methane absorption features. Rather, water absorption features dominate the spectrum (Figure 9). As compared to giant exoplanets having $H_2$-dominated atmospheres, super-Earth exoplanets having $H_2O$-dominated atmospheres would likely be brighter in reflection.

Finally, we summarize all computed models on a series of color-color diagrams for five photometric bands of ~10% band width at 400 – 1000 nm (Figure 10).

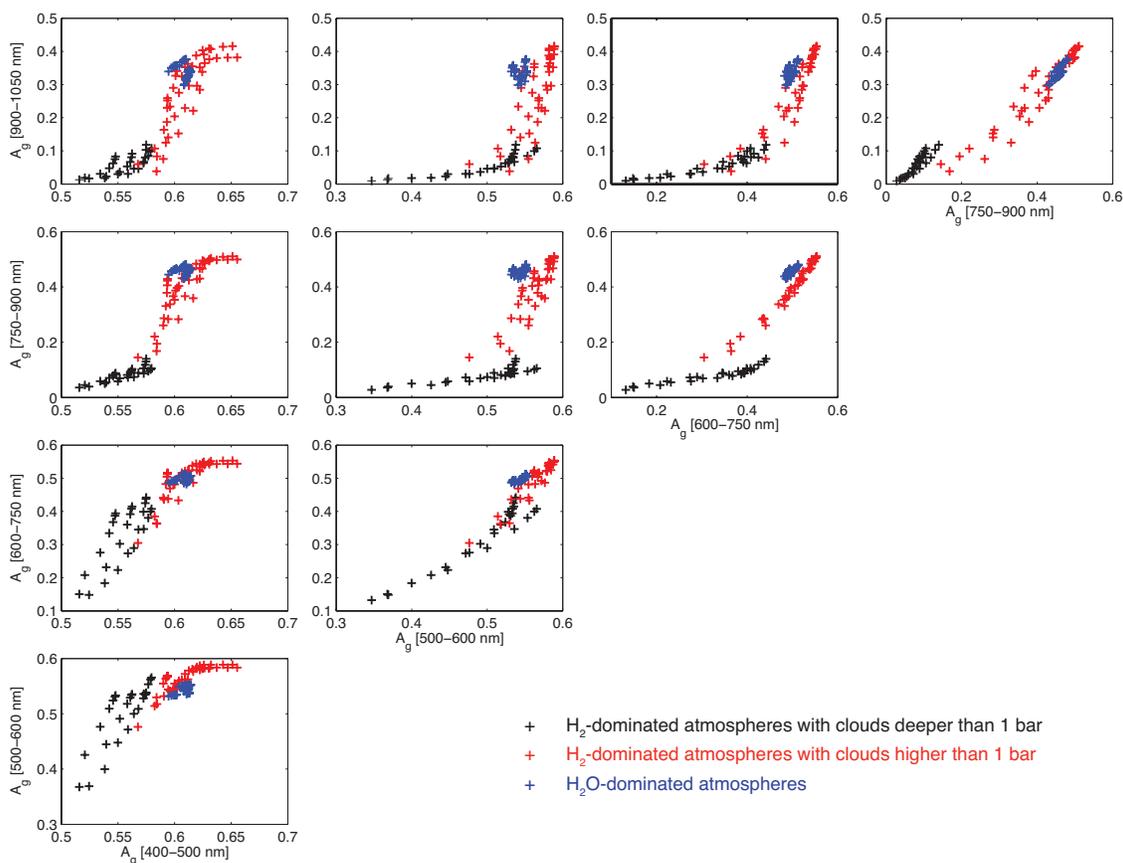

Figure 10: Broadband reflectivity of giant exoplanets having $H_2$-dominated atmospheres and super-Earth exoplanets having $H_2O$-dominated atmospheres, shown as color-color diagrams. The models cover a wide range of plausible planetary scenarios, ranging from $H_2$-dominated atmospheres and $H_2O$-dominated atmospheres, with a semi-major axis ranging from 2.8 to 3.8 AU, a mixing ratio of methane ranging from $10^{-5}$ to $10^{-2}$, an internal heat flux ranging from that of Earth to that of Jupiter, and with and without an photochemical haze layer. The clouds in these models are located from 4 bars to $10^{-3}$ bars.

Because of the muted methane features, super-Earth exoplanets having $H_2O$-dominated atmospheres would occupy a much more confined phase space of the color-color diagrams than giant exoplanets having $H_2$-dominated atmospheres (Figure 10). The giant exoplanets can be grouped into two clearly separated branches: one is the planets having a water cloud deeper than 1 bar as the uppermost cloud deck (black points in Figure 10), and the other is the planets having an ammonia cloud shallower than 1 bar as the uppermost cloud deck (red points in Figure 10). The water-cloud branch features lower geometric albedo at wavelengths longer than 750 nm than the ammonia-cloud branch. For both branches the geometric albedo at the 400 – 500 nm band and the 500 – 600 band is high. There may exist a red haze in the atmosphere that lowers the geometric albedo in these bands. The $H_2O$-rich super-Earth exoplanets having high water clouds overlap with the ammonia-cloud branch of giant planes, but occupy a small phase space on the color-color diagrams. These results show that broadband detections may broadly distinguish $H_2$-rich giant exoplanets versus $H_2O$-rich super-Earth exoplanets, amid ambiguities that may be addressed with modest resolution spectra.

**Acknowledgement**

Renyu Hu gratefully acknowledge the Hubble Post-Doctoral Fellowship Program for supporting this study, and NASA's Exoplanet Exploration Program for requesting it on behalf of the WFIRST/AFTA Science Definition Team and the Exo-S and Exo-C Science and Technology Definition Teams.

**References**

Ackerman, A., Marley, M. S., 2001, ApJ, 556, 872
Atreya, S. K., Wong, M. H., Owen, T. C. et al., 1999, P&SS, 47, 1243
Banfield, D., Gierasch, P. J., Bell, M., et al., 1998, Icarus, 135, 230
Baraffe, I., Chabrier, G., Barman, T., et al., 2003, A&A, 402, 701
Baraffe, I., Chabrier, G., Barman, T., 2008, A&A, 482, 315
Batalha, N. M., 2014, PNAS, 111, 12647
Burrows, A. S., 2014, PNAS, 111, 12601
Burrows, A., Marley, M., Hubbard, W. B., et al., 1997, ApJ, 491, 856
Burrows, A., Sudarsky, D., Hubeny, I., 2004, ApJ, 609, 407
Cahoy, K. L., Marley, M. S., Fortney, J. J., 2010, ApJ, 724, 189
Dlugach, J. M., Yanovitskij, E. G., 1973, Icarus, 22, 66
Elkins-Tanton, L. T., Seager, S., 2008, ApJ, 685, 1237
Fortney, J. J., Saumon, D., Marley, M. S., et al., 2006, ApJ, 642, 495
Gladstone, G. R., Allen, M., Yung, Y. L., 1996, Icarus, 119, 1
Howard, A. W., 2013, Science, 340, 572
Hu, R., Seager, S., Bains, W., 2012, ApJ, 761, 166
Hu, R., Seager, S., Bains, W., 2013, ApJ, 769, 6
Hu, R., Seager, S., 2014, ApJ, 784, 63

Ingersoll, A. P., Dowling, T. E., Gierasch, P. J., et al., 2004, Dynamics of Jupiter's Atmosphere, in Jupiter – The Planet, Satellites and Magnetosphere, F. Bagenal, T. Dowling and W. McKinnon, Eds., Cambridge University Press
Karkoschka, E., 1994, Icarus, 111, 174
Karkoschka, E., 1998, Icarus, 133, 134
Line, M. R., Liang, M. C., Yung, Y. L., 2010, ApJ, 717, 496
Marcy, G. W., Weiss, L. M., Petigura, E. A., et al., 2014, PNAS, 111, 12655
Marley, M. S., Gelino, C., Stephens, D., et al., 1999, ApJ, 513, 879
Matcheva, K. I., Conrath, B. J., Gierasch, P. J., Flasar, F. M., 2005, Icarus, 179, 432
Moses, J. I., Visscher, C., Fortney, J. J., et al., 2011, ApJ, 737, 15
Moses, J. I., Line, M. R., Visscher, C., et al., 2013, ApJ, 777, 34
Sato, M., Hansen, J. E., 1979, JAS, 36, 1133
Seager, S., Sasselov, D., 1998, ApJL, 502, L157
Seager, S., Deming, D., 2010, ARAA, 48, 631
Seager, S., Whitney, B. A., Sasselov, D. D., 2000, ApJ, 540, 504
Seager, S., Turnbull, M., Sparks, W., et al., 2014, Exo-S: Starshade Probe-Class Exoplanet Direct Imaging Mission Concept – Interim Report.
Seiff, A., Kirk, D. B., Knight, T. C. D., et al. 1998, JGR, 103, 22857
Simon-Miller, A., Conrath, B. J., Gierasch, P. J., et al. 2006, Icarus, 180, 98
Spergel, D., Gehrels, N., Breckinridge, J., et al., 2013, Wide-Field InfraRed Survey Telescope-Astrophysics Focused Telescope Assets WFIRST-AFTA Final Report. ArXiv e-prints.
Stapelfeldt, K., Belikov, R., Bryden, G., et al., 2014, Exoplanet Direct Imaging: Coronagraph Probe Mission Study "Exo-C" – Interim Report.
Sudarsky, D., Burrows, A., Pinto, P. 2000, ApJ, 538, 885
Sudarsky, D., Burrows, A., Hubeny, I., 2003, ApJ, 588, 1121
Toon, O. B., McKay, C. P., Ackerman, T. P., 1989, JGR, 94, 16287
Venot, O., Agundez, M., Selsis, F., et al., 2014, A&A, 562, A51
Weidenschilling, S. J., Lewis, J. S., 1973, Icarus, 20, 465
West, R. A., Strobel, D. F., Tomasko, M. G., 1986, Icarus, 65, 161
West, R. A., Orton, G. S., Draine, B. T., Hubbell, E. A., 1989, Icarus, 80, 220
West, R. A., Baines, K. H., Friedson, A. J., et al., 2004, Jovian Clouds and Haze, in Jupiter – The Planet, Satellites and Magnetosphere, F. Bagenal, T. Dowling and W. McKinnon, Eds., Cambridge University Press
Wong, A.-S., Lee, A. Y. T., Yung, Y. L., Ajello, J. M., 2000, ApJ, 534, L215
Wong, A.-S., Yung, Y. L., Friedson, A. J., 2003, GRL, 30, 1447
Yung, Y. L., Strobel, D. F., 1980, ApJ, 239, 395